\documentclass[aps,prl,preprint,showpacs,groupedaddress]{revtex4}
\usepackage {amssymb}
\usepackage {amsmath}
\usepackage {graphicx}
\usepackage {longtable}

\begin{document}
\title{The dissolution of the vacancy gas in solids and high pressure phase
transitions}

\author{Fedor V.Prigara}
\affiliation{Institute of Microelectronics and Informatics,
Russian Academy of Sciences,\\ 21 Universitetskaya, Yaroslavl
150007, Russia} \email{fvprigara@rambler.ru}

\date{\today}

\begin{abstract}

The formula for the number density of vacancies in a solid under pressure is
obtained. The mean number density of vacancies in a solid under stress or
tension is also estimated. The dissolution of the vacancy gas in solids is
shown to be responsible for the phases with composite incommensurate
structures in metals under high pressure and also for the low values of the
elastic limit and tensile strength of solids.

\end{abstract}


\maketitle

In the last decade, the phases with composite incommensurate structures have
been discovered in metals and some other elemental solids under high
pressure [1-3] (for a review see Ref. 4). Similar phenomenon is the decrease
of the melting temperature of sodium under pressure [5]. All these effects
can be explained by the dissolution of the vacancy gas in solids under
pressure following from the thermodynamic consideration.

According to the thermodynamic equation [6]

\begin{equation}
\label{eq1}
dE = TdS - pdV,
\end{equation}

\noindent
where \textit{E} is the energy, \textit{T} is the temperature, \textit{S} is
the entropy, \textit{p} is the pressure, and \textit{V} is the volume of a
solid, the energy of a solid increases with pressure, so the pressure acts
as the energy factor similarly to the temperature. Therefore, the number of
vacancies in a solid increases both with temperature and with pressure.

Consider at first the dissolution of the vacancy gas in solids at ambient
pressure. We treat the dissolution of vacancies in a solid as a phase
transition (the evaporation of vacancies from the `condensed phase' at the
surface of the solid) and apply the Clausius- Clapeyron equation [6] in the
form

\begin{equation}
\label{eq2}
\frac{{dP}}{{dT}} = \frac{{E_{v}} }{{Tv}},
\end{equation}

\noindent
where $E_{v} $ is the energy of the vacancy formation, \textit{v} is the
volume per one vacancy, and the pressure of the vacancy gas in a solid is
given by the formula

\begin{equation}
\label{eq3}
P = nT = T/v,
\end{equation}

\noindent
where \textit{n} is the number density of vacancies. Here the Boltzmann
constant $k_{B} $ is included in the definition of the temperature
\textit{T}.

From equations (\ref{eq2}) and (\ref{eq3}) we obtain

\begin{equation}
\label{eq4}
n = \left( {P_{0} /T} \right)exp\left( { - E_{v} /T} \right) = \left( {n_{0}
T_{0} /T} \right)exp\left( { - E_{v} /T} \right),
\end{equation}

\noindent
where $P_{0} = n_{0} T_{0} $ is a constant, $T_{0} $ can be put equal to the
melting temperature of the solid at ambient pressure, and the constant
$n_{0} $ has an order of magnitude of the number density of atoms in the
solid.

The obtained formula (\ref{eq4}) describes the thermal expansion of the solid. It
should be taken into account that the dissolution of the vacancy gas in a
solid causes the deformation of the crystalline lattice and changes the
lattice parameters.

The energy of the vacancy formation depends on the temperature and
pressure, the first terms of the corresponding series expansion
having the form

\begin{equation}
\label{eq5}
E_{v} = E_{0} - cT - \alpha P/n_{0} ,
\end{equation}

\noindent where \textit{c} and $\alpha $ are dimensionless
constants. The temperature dependence of $E_{v} $ in the equation
(\ref{eq5}) causes only the change of a constant $P_{0} $ in the
equation (\ref{eq4}). The effect of the pressure dependence of the
energy of the vacancy formation is much more essential. On the
atomic scale, the pressure dependence of the energy of the vacancy
formation in the equation (\ref{eq5}) is produced by the strong
atomic relaxation in a crystalline solid under high pressure. Note
that we do not consider here the comparatively narrow region of
small pressures nearby the melting curve, where the sign of the
coefficient $ \alpha$ is negative [7].

With increasing pressure, the number density of vacancies in a solid
increases, according to the relation

\begin{equation}
\label{eq6}
n = \left( {n_{0} T_{0} /T} \right)exp\left( { - \left( {E_{0} - \alpha
P/n_{0}}  \right)/T} \right),
\end{equation}

\noindent
and, finally, the vacancies can condense, forming their own sub-lattice.
Such is the explanation of the appearance of composite incommensurate
structures in metals and some other elemental solids under high pressure
[1-4]. For example, the structure of phase IV of barium is composed of a
tetragonal `host', with `guest' chains in channels along the \textit{c} axis
of the host . These chains form two different structures, one well
crystallized and the other highly disordered. The guest structures are
incommensurate with the host [1]. Rb- IV has an incommensurate composite
structure, comprising a tetragonal host framework and a simple body-
centered tetragonal guest, the ratio of the \textit{c}-axis lattice
parameters being strongly pressure dependent [2].

Further increase of the number density of vacancies in a solid
with increasing pressure leads to the melting of the solid under
sufficiently high pressure (and fixed temperature). Such effect
has been observed in sodium [5]. In general, such behavior is
universal for solids, though the corresponding melting pressure is
typically much larger than those for sodium. The melting pressure
$P_{m}$ at low temperatures can be estimated from the relation

\begin{equation}
\label{eq7}
P_{m} \cong n_{0} T_{m} ,
\end{equation}

\noindent
where $T_{m} $ is the melting temperature at ambient pressure. The
temperature dependence of the melting pressure (or, equivalently, the
pressure dependence of the melting temperature in the region of high
pressures) is considered below.

At the melting pressure $P = P_{m} $, the energy of the vacancy
formation (5) is small and can be put to be approximately equal to
zero,

\begin{equation}
\label{eq8} E_{v} = E_{0} - \alpha P/n_{0} \approx 0.
\end{equation}

Taking into account the relation (7), from the equation
(\ref{eq8}) we obtain

\begin{equation}
\label{eq9} \alpha \approx E_{0} n_{0} /P_{m} \cong E_{0} /T_{m} .
\end{equation}

There is an empirical relation between the activation energy of
self diffusion and the melting temperature of a solid [7]. Due to
the atomic relaxation in a crystalline solid, the migration
barrier for a vacancy is low, so that the activation energy of
self- diffusion is approximately equal to the energy of the
vacancy formation. For the most of metals, this relation has a
form

\begin{equation}
\label{eq10} E_{0} \approx 18T_{m} ,
\end{equation}

\noindent and from the equation (\ref{eq9}) we find $\alpha
\approx 18$. Thus, for the most of metals, the formula

\begin{equation}
\label{eq11} n \cong \left( {n_{0} T_{0} /T} \right)exp\left( { -
\left( {E_{0} - 18P/n_{0}}  \right)/T} \right)
\end{equation}

\noindent is valid. For anomalous body centered cubic metals ($Ti
- \beta $, $Zr - \beta $, $U - \gamma $, and others), at low
temperatures the relation $E_{0} \approx 12T_{m} $ is valid, so in
this case $\alpha \approx 12$.

The low values of the activation energy of self-diffusion in
anomalous \textit{bcc} metals seem to be produced by the stresses
associated with the phase transition from a low-temperature dense
phase to a high-temperature \textit{bcc} phase [7]. The effect has
a local character in the temperature- pressure plane, so in the
region of high pressures the value of the coefficient alpha,
$\alpha \approx 18$, is universal for crystalline solids.

Replacing in the relation (\ref{eq6}) the pressure \textit{P} by
the absolute value of the stress or tension $\sigma = F/S$,
applied to a solid, where \textit{F} is the applied force and
\textit{S} is the cross-section area of the solid in the plane
perpendicular to the direction of the applied force, we can
estimate the mean number density of vacancies in the solid under
the stress or tension:

\begin{equation}
\label{eq12} \langle n\rangle \cong \left( {n_{0} T_{0} /T}
\right)exp\left( { - \left( {E_{0} - \alpha \sigma /n_{0}}
\right)/T} \right).
\end{equation}

Here, for the most of metals the value of the constant $\alpha $
is $\alpha \approx 18$, and for the anomalous \textit{bcc} metals
this constant is $\alpha \approx 12$.

The dissolution of the vacancy gas in a solid under the stress or
tension is responsible for the low values of the elastic limit and
the tensile strength of solids as compared with theoretical
estimations not taking into account this process [8].

As can be inferred from the relation (\ref{eq7}) as compared with
the experimental data, there are some systematic errors in the
calibration of high pressure. This conclusion is supported by
those fact that the ratio of `measured' pressure and the value of
pressure expected from the relation (\ref{eq7}) has the same order
of magnitude as the ratio of theoretical (calculated) and real
tensile strengths of solids [8].

The melting pressure at low temperatures has an order of magnitude of the
tensile strength of a solid, $P_{m} \cong \sigma _{s} $, so from equation
(\ref{eq7}) we find

\begin{equation}
\label{eq13} n_{0} \cong \sigma _{s} /T_{m} .
\end{equation}

This relation gives another estimation of the constant $n_{0} $.
For iron $T_{m} = 1800K = 1.88 \times 10^{ - 13}erg$, $\sigma _{s}
= 3 \times 10^{9}dyn/cm^{2}$, and relation (\ref{eq13}) gives
$n_{0} \cong 1.2 \times 10^{22}cm^{ - 3}$. For silver $T_{m} =
1230K = 1.7 \times 10^{ - 13}erg$, $\sigma _{s} = 1.8 \times
10^{9}dyn/cm^{2}$, and equation (\ref{eq13}) gives $n_{0} \cong
1.1 \times 10^{22}cm^{ - 3}$.

Since the pressure acts as an energy factor similar to the
temperature, it normally causes the decrease of the critical
temperature of phase transitions in solids. For example, for
single- metal superconductors the critical temperature of the
phase transition is found to decrease under pressure [9].
Ferromagnets also normally show the decrease of the Curie
temperature with pressure [10]. Similar is the behavior of the
Curie temperature of ferroelectrics [11].

Superconducting phase transitions in metals and cuprate oxides
under high pressure [9,12,13] are described by the relation

\begin{equation}
\label{eq14} T_{c} + kP/n_{0} = const,
\end{equation}

\noindent where $T_{c} $ is the critical temperature of the
transition and \textit{k} is a constant dependent on the chemical
composition. The relation (\ref{eq14}) is also valid for weak
itinerant ferromagnets [10] and ferroelectrics [11]. The equation
(\ref{eq14}) describes also the melting of a solid under
sufficiently high pressure, the constant \textit{k} in this case
being $k \approx 1$ [14].

To summerize, the dissolution of the vacancy gas in solids under
pressure determines both the structural transitions and the
melting of a solid in the high- pressure region. The dissolution
of vacancies in solids under the stress or tension leads to the
low values of the elastic limit and the tensile (or compression)
strength of solids. The treating of the pressure as an energy
factor similar to the temperature can explain the linear
dependence of the critical temperature of a phase transition on
the pressure observed in superconductors, ferromagnets, and
ferroelectrics.

The author is grateful to V.P.Prigara for useful discussion.

\begin{center}
---------------------------------------------------------------
\end{center}

[1] R.J.Nelmes, D.R.Allan, M.I.McMahon, and S.A.Belmonte, Phys. Rev. Lett.
\textbf{83}, 4081 (1999).

[2] M.I.McMahon, S.Rekhi, and R.J.Nelmes, Phys. Rev. Lett. \textbf{87},
055501 (2001).

[3] O.Degtyareva, E.Gregoryanz, M.Somayazulu, H.K.Mao, and R.J.Hemley, Phys.
Rev. B \textbf{71}, 214104 (2005).

[4] V.F.Degtyareva, Usp. Fiz. Nauk \textbf{176}, 383 (2006)
[Physics- Uspekhi \textbf{49} (2006)].

[5] E.Gregoryanz, O.Degtyareva, M.Somayazulu, R.J.Hemley, and
H.K.Mao, Phys. Rev. Lett. \textbf{94}, 185502 (2005).

[6] S.-K.Ma, \textit{Statistical Mechanics} (World Scientific, Philadelphia,
1985).

[7] B.S.Bokstein, S.Z.Bokstein, and A.A.Zhukhovitsky,
\textit{Thermodynamics and Kinetics of Diffusion in Solids}
(Metallurgy Publishers, Moscow, 1974).

[8] G.I.Epifanov, \textit{Solid State Physics} (Higher School
Publishers, Moscow, 1977).

[9] S.Deemyad and J.S.Schilling, Phys. Rev. Lett. \textbf{91},
167001 (2003).

[10] J.J.Hamlin, S.Deemyad, J.S.Schilling et al., E-print
archives, cond-mat/0609149.

[11] G.S.Zhdanov, \textit{Solid State Physics} (Moscow University
Press, Moscow, 1961).

[12] J.J.Hamlin, V.G.Tissen, and J.S.Schilling, Phys. Rev. B
\textbf{73}, 094522 (2006).

[13] M.K.Crawford, R.L.Harlow, S.Deemyad et al., Phys. Rev. B
\textbf{71}, 104513 (2005).

[14] F.V.Prigara, \textit{The Structure of Earth's Crust}
(unpublished).

\end{document}